\documentclass[aps,prx,reprint,groupedaddress]{revtex4-1}
\usepackage{amsmath,amssymb,mathtools}
\usepackage{graphicx}
\usepackage{dcolumn}
\usepackage{bm}
\usepackage[hidelinks]{hyperref}
\hypersetup{pdfpagemode=UseNone}
\usepackage{siunitx}
\usepackage{physics2}
\usephysicsmodule{braket}

\newcommand{\sSz}{{}^1\mathrm{S}_0}
\newcommand{\sPo}{{}^1\mathrm{P}_1}

\newcommand{\tPo}{{}^3\mathrm{P}_1}
\newcommand{\tPt}{{}^3\mathrm{P}_2}
\newcommand{\tSo}{{}^3\mathrm{S}_1}
\newcommand{\boson}{{}^{174}\mathrm{Yb}}

\begin{document}

\title{Development of a high-power ultraviolet laser system and observation of fast coherent Rydberg excitation of ytterbium}

\author{Yuma Nakamura}\email[]{nakamura.yuma.54c@st.kyoto-u.ac.jp}
 \affiliation{Department of Physics, Graduate School of Science, Kyoto University, Kyoto 606-8502, Japan}
 \author{Naoya Ozawa}
  \affiliation{Department of Physics, Graduate School of Science, Kyoto University, Kyoto 606-8502, Japan}
\author{Toshi Kusano}
  \affiliation{Department of Physics, Graduate School of Science, Kyoto University, Kyoto 606-8502, Japan}
\author{Rei Yokoyama}
  \affiliation{Department of Physics, Graduate School of Science, Kyoto University, Kyoto 606-8502, Japan}
\author{Kosuke Shibata}
  \affiliation{Department of Physics, Graduate School of Science, Kyoto University, Kyoto 606-8502, Japan}
\author{Tetsushi Takano}
  \affiliation{Department of Physics, Graduate School of Science, Kyoto University, Kyoto 606-8502, Japan}
\author{Yosuke Takasu}
  \affiliation{Department of Physics, Graduate School of Science, Kyoto University, Kyoto 606-8502, Japan}
\author{Yoshiro Takahashi}
  \affiliation{Department of Physics, Graduate School of Science, Kyoto University, Kyoto 606-8502, Japan}
  
\date{\today}

\begin{abstract}
We present the development of a high-power ultraviolet laser system operating at a wavelength of \qty{325}{nm} for Rydberg excitation from the $\tPt$ state of ytterbium. Utilizing a two-stage frequency doubling scheme, we achieved an output power exceeding \qty{800}{mW}. The system effectively suppresses frequency noise in the MHz range, which is critical for achieving high Rydberg excitation fidelity, through the use of a filtering cavity. Using this system, we demonstrated coherent excitation of the $(6s71s)\tSo$ Rydberg state with a Rabi frequency of \qty{2.13(3)}{MHz}. Combined with our successful manipulations on the $\sSz\text{--}\tPt$ transition, this work represents a foundational step toward achieving high-fidelity Rydberg excitation, enabling advancements in quantum simulation and computing with neutral atom arrays.
\end{abstract}

\maketitle

\section{Introduction}
\label{sec:introduction} 
Rydberg interactions in neutral atom arrays play crucial roles in quantum simulations of various physical models \cite{Labuhn2016-ie,Bernien2017-ab,Lienhard2018-rl,de-Leseleuc2019-aw,Semeghini2021-vn,Scholl2022-fn,Scholl2021-dr,Ebadi2021-jb,Gonzalez-Cuadra2022-in} and in generating entanglements for quantum information processing \cite{Jaksch2000-su,Lukin2001-wv,Jandura2022-ei,Isenhower2010-jw,Levine2018-ge,Levine2019-os,Evered2023-nv}. The Rydberg states, which are highly excited states close to the ionization limit of an atom, are typically accessed via a two-photon process in alkali atoms such as rubidium or cesium. However, this method can introduce unwanted scattering from the intermediate state, which can negatively impact the Rydberg excitation fidelity\cite{Saffman2016-jh}.

In contrast, alkali-earth or alkali-earth-like atoms (AEA) such as strontium \cite{Cooper2018-sk,Norcia2018-rm,Barnes2022-wy,Urech2022-wd,Tao2024-ty,Unnikrishnan2024-ec,Yan2024-qi} and ytterbium \cite{Saskin2019-xk,Okuno2022-cd,Jenkins2022-lt,Huie2023-wp,Norcia2023-dt} have recently attracted considerable interest. The optical tweezer systems employing the AEA enable Rydberg excitation via a single-photon process from a metastable excited state, offering a pathway to higher-fidelity operations \cite{Madjarov2020-kz,Schine2022-sn,Scholl2023-jb,Cao2024-hp,Tsai2024-tu,Peper2024-wc}. The key to further improving fidelity in these setups lies in the use of high-power lasers with low frequency noise \cite{Jiang2023-dp,Tsai2024-tu}. However, this introduces a technical challenge: the wavelength required for Rydberg excitation in AEAs lies in the ultraviolet (UV) region, making it inherently difficult to construct a high-power laser with the necessary specifications for such systems.

Here, we report on the development of a high-power UV laser system with frequency noise suppression for Rydberg excitation from the metastable $\tPt$ state of ytterbium (Yb). Note that the two-level system consisting of the long-lived $\tPt$ and ground $\sSz$ states offers the novel possibility of an optical clock qubit with non-destructive measurement capability \cite{Nakamura2024-fz}. We successfully obtained more than \qty{800}{mW} output at a wavelength of \qty{325}{nm} utilizing a Raman fiber amplifier (RFA) and intracavity second harmonic generation (SHG). The frequency noise in the MHz range is effectively suppressed by \qty{20}{dB} using a filtering cavity. We employed this laser system for Rydberg spectroscopy and coherent excitation of single $\boson$ atoms in an optical tweezer array. We successfully observed the Rabi oscillation between the $(6s6p)\tPt$ and $(6s71s){}^3\mathrm{S}_1$ Rydberg states with a frequency of \qty{2.13(3)}{MHz}. Combined with our successful observation of coherent excitation of the $\sSz\text{--}\tPt$ optical clock qubit with Rabi oscillation and Ramsey signal, this work is the first step for high-fidelity Rydberg excitation to facilitate quantum simulation and information processing in neutral atom arrays.
\begin{figure*}[t]
    \centering
    \includegraphics[width=\textwidth]{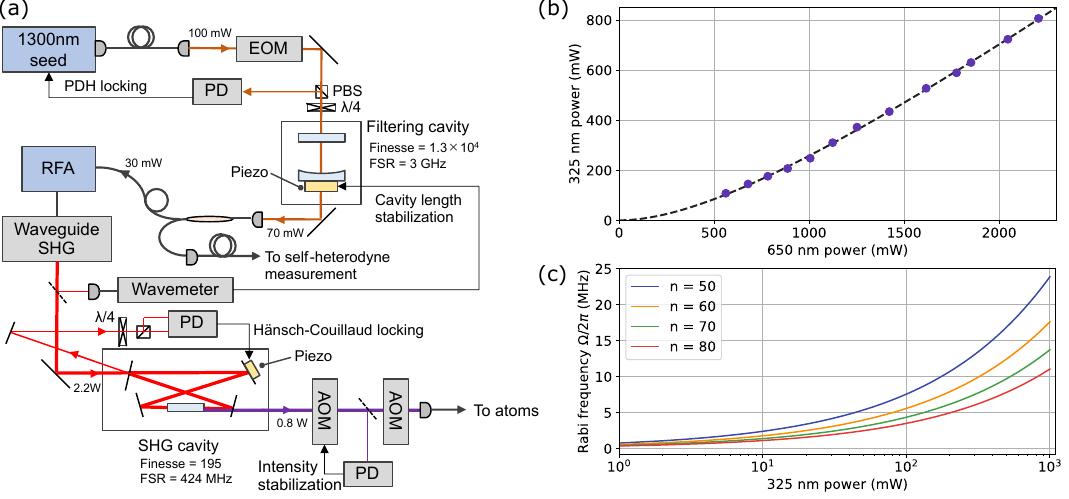}
    \caption{The developed laser system. (a) Overview of the laser system. The \qty{1300}{nm} seed laser of \qty{100}{mW} is output from a single-mode fiber and fed into the filtering cavity after passing through an EOM for PDH locking. The cavity length is stabilized via a wavemeter. The transmitted light with a power of \qty{70}{mW} is then coupled to a fiber and divided by a $1\times2$ fiber splitter. One arm of the splitter is used for delayed self-heterodyne measurements and the other is sent to the RFA. The output from the RFA is converted to \qty{650}{nm} light of \qty{2.2}{W} by the waveguide SHG module and fed into the SHG cavity system we developed, which is locked by the H\"{a}nsch-Couillaud method. The \qty{325}{nm} beam from the SHG cavity passes through two AOMs, one for intensity stabilization and the other for pulse shaping, and couples to a single-mode fiber to the experimental chamber. Finally, the beam is focused onto the atoms with a beam radius of \qty{70}{\mu m}. PD represents a photo-diode. (b) The \qty{325}{nm} output power after the SHG cavity we developed. The maximum power exceeds \qty{800}{mW}. The dashed line is the fitting curve with Eqs.\,(\ref{eq:SHG_formulae}) and (\ref{eq:intracavity_power}) with the fitting parameter of $l_{\mathrm{cav}}$. With the independently measured values of $E_{\mathrm{nl}}=9.4(1)\times10^{-5}\,\mathrm{/W}$, $T_1=\qty{2.1}{\%}$, and $\eta=\qty{82}{\%}$, we estimate $l_{\mathrm{cav}}=\qty{1.08(8)}{\%}$, consistent with the cavity finesse of 195. (c) Expected Rabi frequencies with various \qty{325}{nm} beam powers. In the calculation, the reduced dipole matrix elements between $6p$ and $ns$ states are taken from previous research \cite{Covey2019-sr}, and a beam radius of \qty{70}{\mu m} at the atom position is assumed.}
    \label{fig:figure1}
\end{figure*}

\section{A high-power UV laser system}
\subsection{Overview of the system}
The transition wavelength from the metastable $\tPt$ state to a Rydberg state in Yb is approximately \qty{325}{nm} \cite{Okuno2022-cd}. To achieve high-power output at this wavelength, we employ a two-stage frequency doubling scheme (Fig.\,\ref{fig:figure1}(a)). The laser system is composed of a seed laser ($\lambda = \qty{1300}{nm}$, Toptica, TA pro), followed by a Raman fiber amplifier coupled with a waveguide SHG module (MPB Communications, VRFA-SF), which converts the wavelength from \qty{1300}{nm} to \qty{650}{nm}. The second stage of frequency doubling is achieved by developing an efficient SHG cavity system utilizing a beta barium borate (BBO) crystal, which further converts the wavelength from \qty{650}{nm} to \qty{325}{nm}.

The seed laser, with an output power of 100 mW, is first coupled to a single-mode fiber. It is then passed through an electro-optic modulator (EOM) before being fed into a Fabry-P\'{e}rot cavity for frequency stabilization using the Pound-Drever-Hall (PDH) locking technique \cite{Drever1983-ke}. The cavity also serves as a filter to suppress the frequency noise to improve Rydberg excitation fidelity \cite{Levine2018-ge}. The cavity used in this setup has a finesse of $1.3\times10^4$ and a free spectral range (FSR) of \qty{3}{GHz}, resulting in a \qty{20}{dB} suppression in frequency noise of the transmitted light at an offset frequency of 1.5 MHz, as discussed later. The cavity length is tunable by piezoelectric transducers attached between the two mirror-holder tubes and locked to the desired frequency using a wavemeter (HighFinesse, WS8-10). This tunability enables us to lock the seed laser to various Rydberg resonances, optimizing the system for specific quantum operations.

The transmitted light from the filtering cavity, with an output power of \qty{70}{mW}, is coupled into a single-mode fiber and subsequently split using a $1\times2$ fiber splitter with a $50:50$ ratio. One arm of the splitter, delivering approximately \qty{30}{mW} of power, is directed to the RFA, while the other arm is employed in a self-heterodyne measurement for frequency noise evaluation. The \qty{1300}{nm} light, amplified by the RFA, is then converted to \qty{2.5}{W} of \qty{650}{nm} light via the waveguide SHG module.

The converted \qty{650}{nm} light is sent to the SHG cavity we developed, which consists of four mirrors (three mirrors with a reflectivity $R>\qty{99.8}{\%}$ and an input coupler with a transmittance $T_1=\qty{2.1}{\%}$ at \qty{650}{nm}, and all mirrors $R<\qty{2}{\%}$ at \qty{325}{nm}) and a BBO crystal (\qty{3}{mm}(height)$\times$\qty{3}{mm}(width)$\times$\qty{10}{mm}(length), with cut angles of $\theta=36.7^\circ$, $\phi=90.0^\circ$). The nonlinear coefficient of the BBO crystal $E_{\mathrm{nl}}$ is $9.4(1)\times10^{-5}\,\mathrm{/W}$, measured by a single-pass SHG setup.
The cavity finesse and FSR are 195 and \qty{424}{MHz}, respectively. Figure \ref{fig:figure1}\,(b) shows the output power of the SHG cavity under H\"{a}nsch-Couillaud locking \cite{Hansch1980-nr}. We obtain more than \qty{800}{mW} output, which is sufficient for fast Rydberg excitation at MHz-order  (Fig.\,\ref{fig:figure1}\,(c)). By fitting this result with the theoretical SHG output power $P_{\mathrm{SHG}}$ given as \cite{Polzik1991-oh,Pizzocaro2014-ep}
\begin{equation}
    P_{\mathrm{SHG}} = E_{\mathrm{nl}}P_{\mathrm{ic}}^2,
    \label{eq:SHG_formulae}
\end{equation}
with 
\begin{equation}
    P_{\mathrm{ic}} = \frac{\eta T_1 P_{\mathrm{in}}}{(1-\sqrt{(1-T_1)(1-l_{\mathrm{cav}})(1-E_{\mathrm{nl}}P_{\mathrm{ic}})})^2},
    \label{eq:intracavity_power}
\end{equation}
where $P_{\mathrm{ic}}$ is the intracavity power of \qty{650}{nm} light, $l_{\mathrm{cav}}$ is the round-trip power loss in the cavity except for the input coupler, $\eta$ is the coupling efficiency to the cavity, and $P_{\mathrm{in}}$ is the input power to the cavity, we determine the cavity loss as $l_{\mathrm{cav}}=\qty{1.08(8)}{\%}$, with the independently measured value of $\eta = \qty{82}{\%}$. The estimated $l_{\mathrm{cav}}$ is consistent with the cavity finesse of 195.

The output from the SHG cavity passes through two acousto-optic modulators (AOMs), one for intensity stabilization \cite{Tricot2018-cx} and the other for fast pulse shaping. Following this, the light is delivered to the experimental chamber via a single-mode fiber and focused onto the atoms trapped in an optical tweezer array, with a beam radius of approximately \qty{70}{\mu m}. We note that the total power at the atom position is roughly \qty{20}{\%} of the SHG cavity output, which could be further increased by optimizing the spatial beam mode to better match the fiber mode.

\subsection{Characteristics of the laser noises}\label{sec:laser_noise}

\begin{figure}[t!]
    \centering
    \includegraphics[width=0.98\linewidth]{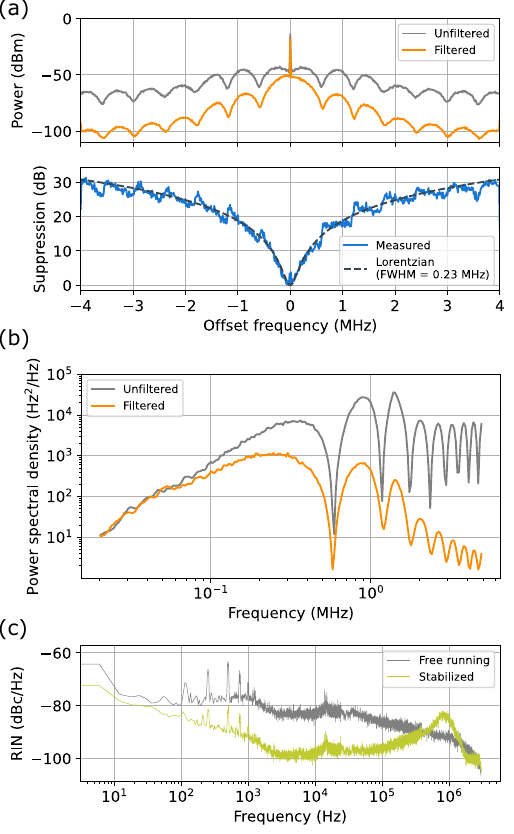}
    \caption{Characteristics of the laser noises. (a) (Top) The spectra of the delayed self-heterodyne beat notes of the seed laser with and without the filtering cavity. The periodic dip structure corresponds to the delay line length of \qty{350}{m} \cite{Jiang2023-dp}. (Bottom) The suppression factor calculated from the ratio between the spectra. As a guide for the eye, a Lorentzian function with a full width at half maximum of \qty{230}{kHz}, corresponding to the filtering cavity linewidth, is plotted with a dashed line. (b) The frequency noise PSDs of the unfiltered and filtered seed laser, evaluated from the delayed self-heterodyne signals. The frequency noise including the servo bump around \qty{1}{MHz} of the unfiltered seed laser is effectively suppressed by the filtering cavity. The measurement sensitivity decreases at multiples of the Fourier frequency equivalent to the inverse of the delay time resulting in the periodic dip structure, as well as at low Fourier frequencies. Thus, the obtained PSD underestimates the actual frequency noise in these regions. (c) The relative intensity noise spectra of \qty{325}{nm} light. The intensity noise below \qty{100}{kHz} is effectively suppressed by the AOM-based stabilization. The increasing noise above \qty{500}{kHz} is due to the limited bandwidth of the AOM, which can be addressed by installing an EOM-based feed-forward stabilization method \cite{Wang2020-mr}.}
    \label{fig:figure2}
\end{figure}

It is crucial to characterize both frequency and intensity noises in the laser to estimate the fidelity of Rydberg excitation \cite{Tsai2024-tu,Jiang2023-dp}. Frequency noise around the Rydberg excitation Rabi frequency, preferably a few MHz and higher \cite{Tsai2024-tu}, negatively impacts the coherent dynamics during excitation \cite{Levine2018-ge} and should therefore be minimized. To achieve this, we employ a filtering cavity, as mentioned earlier. We measure the suppression factor for frequency noise in the MHz range by comparing the delayed self-heterodyne spectra of the seed laser with and without the filtering cavity. The spectra and suppression factor are presented in Fig.\ref{fig:figure2}\,(a). We achieve more than \qty{20}{dB} suppression in the MHz range, consistent with the filtering cavity linewidth (full width at half maximum) of \qty{230}{kHz}. 

Owing to the effective suppression of frequency noise in the MHz range, the power spectral density (PSD) of the frequency noise at \qty{1}{MHz} for the filtered seed laser is estimated to be \qty{7e2}{Hz^2/Hz}, as determined from the phase noise analysis of the delayed self-heterodyne signals using an oscilloscope (Keysight, MXR608A) (Fig.\ref{fig:figure2}\,(b)). According to the previous study \cite{Jiang2023-dp}, and assuming a Rabi frequency of \qty{1}{MHz}, this corresponds to an error rate in Rydberg excitation of approximately $8\times10^{-4}$, which surpasses the threshold required for quantum error correction \cite{Fowler2009-jf}. Further reduction in frequency noise could be achieved by replacing the seed laser with one that exhibits lower noise characteristics.

The relative intensity noise (RIN) spectra of \qty{325}{nm} light, with and without AOM-based intensity stabilization, are shown in Fig.\ref{fig:figure2}\,(c). The RIN at the kHz range, which arises from the limited bandwidth of the SHG cavity locking and the frequency noise of the seed laser, is effectively suppressed by stabilization. However, stabilization becomes ineffective above \qty{500}{kHz} due to the limited bandwidth of the AOM. As the Rabi frequency exceeds \qty{1}{MHz}, the Rydberg excitation error is affected by the RIN integrated up to this range. As a result, the intensity stabilization increases the integrated RIN up to \qty{3}{MHz} to approximately \qty{5}{\%}, slightly larger than that without intensity stabilization. This leads to a Rydberg excitation error of \qty{0.15}{\%} estimated from previous research \cite{Jiang2023-dp}. To address this, enhancing the stabilization bandwidth by incorporating an EOM-based feed-forward stabilization technique \cite{Wang2020-mr} would be effective.

\section{Coherent manipulation on the \textsuperscript{1}S\textsubscript{0}--\textsuperscript{3}P\textsubscript{2} transition}

\begin{figure}[t!]
    \centering
    \includegraphics[width=\linewidth]{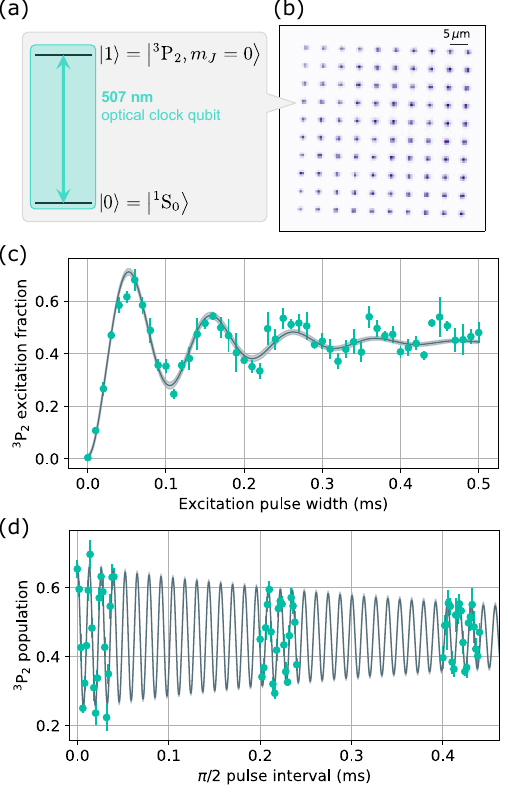}
    \caption{Coherent manipulation on the $\sSz\text{--}\tPt$ transition. (a) The optical clock qubit encoded in the ground and the $m_J=0$ of the $\tPt$ states, offering the capability of non-destructive qubit readout. (b) Averaged fluorescence image of $\boson$ trapped in $10\times10$ optical tweezer array. The site interval is set to \qty{5}{\mu m} in this experiment. (c) Rabi oscillation between the ground and the $\tPt$ states. The gray curve is a fitting function with a damping sine function, yielding a Rabi frequency of $2\pi\times\qty{9.5(2)}{kHz}$. The shaded area is $1\sigma$-confidence intervals of the fitting. The error bars represent the standard error of the mean. (d) The coherence of the optical clock qubit measured by the Ramsey sequence. From the fitting with a damping sine function, the coherence time is estimated to be \qty{0.55(7)}{ms}. The error bars represent the standard error of the mean.}
    \label{fig:figure3}
\end{figure}

Before presenting the experimental results on the Rydberg excitation using the developed laser system, we first demonstrate coherent manipulation of the $\sSz\text{--}\tPt$ transition in $\boson$ atoms, which serves as an optical clock qubit with the inherent capability for non-destructive qubit readout \cite{Nakamura2024-fz} (Fig.\ref{fig:figure3}\,(a)). This $\tPt$ excitation also constitutes the initial step in the Rydberg excitation process.

The loading procedure for single atoms into an optical tweezer array is the same as that used in previous research \cite{Okuno2022-cd,Nakamura2024-fz}. Here, we briefly summarize the sequence. The experiment starts with Zeeman slowing with the $\sSz\text{--}\sPo$ transition ($\lambda=\qty{399}{nm}$), followed by a magneto-optical trapping using the $\sSz\text{--}\tPo$ transition ($\lambda=\qty{556}{nm}$). The atoms are then loaded to a $10\times10$ optical tweezer array generated by a spatial light modulator and a \qty{532}{nm} laser, which is a near-magic wavelength for the $\sSz\text{--}\tPo$ and $\sSz\text{--}\tPt$ transitions of $\boson$ \cite{Yamamoto2016-vk,Saskin2019-xk,Okuno2022-cd}. The trap depth is approximately $k_{\mathrm{B}}\times\qty{0.9}{mK}$ with a tweezer beam radius of $\sim\qty{550}{nm}$, where $k_\mathrm{B}$ is the Boltzmann constant. The site interval is set to $\qty{5}{\mu m}$. After applying a \qty{556}{nm} beam, which induces light-assisted collision between the atoms, to ensure the number of atoms in the tweezers is zero or one, we image the single atoms with an exposure time of \qty{60}{ms} by irradiating a \qty{399}{nm} probe beam with an intensity of $1\times10^{-3}\,I_{s,399}$ and simultaneous \qty{556}{nm} cooling beams with total intensity of $10\,I_{s,556}$ (Fig.\ref{fig:figure3}\,(b)). Here, $I_{s,399}$ and $I_{s,556}$ are the saturation intensities for the $\sSz\text{--}\sPo$ and $\sSz\text{--}\tPo$ transitions, respectively. The survival probability of atoms after imaging is approximately \qty{95}{\%}. After imaging, the atoms are further cooled by the \qty{556}{nm} beams to a temperature of \qty{25}{\mu K}, providing the starting conditions for the $\tPt$ and Rydberg excitation experiments.

To coherently excite the atoms to the $\tPt$ state, we ramp down the tweezer trap to $k_\mathrm{B}\times\qty{35}{\mu K}$ in \qty{10}{ms} to suppress the tweezer-induced-scattering from the $\tPt$ state and apply a magnetic field of \qty{5}{G} along the tweezer polarization in the horizontal plane to make a near-magic condition, where the differential light shift between the $\sSz$ state and the $m_J=0$ state of the $\tPt$ is minimized. We then apply a \qty{507}{nm} beam with vertical linear polarization in the horizontal plane. The beam has a radius of \qty{105}{\mu m} (horizontal) and \qty{45}{\mu m} (vertical) with a power of \qty{210}{mW}.  Figure \ref{fig:figure3}\,(c) illustrates the Rabi oscillation between the ground state and the $m_J=0$ state of the $\tPt$, exhibiting a Rabi frequency of $2\pi \times \qty{9.5(2)}{kHz}$. It should be noted that the Rabi oscillation is primarily damped by the thermal distribution of the motional states in the tweezer potential \cite{Takamoto2009-rp}. In this experiment, the mean vibrational quantum number $\bar{n}$ is approximately $\bar{n}=1$, resulting in variations in the Rabi frequency due to motional coupling. This can be addressed through sideband cooling to the motional ground state via the clock transition \cite{Zhang2022-gp}.
\begin{figure*}[t!]
    \centering
    \includegraphics[width=\textwidth]{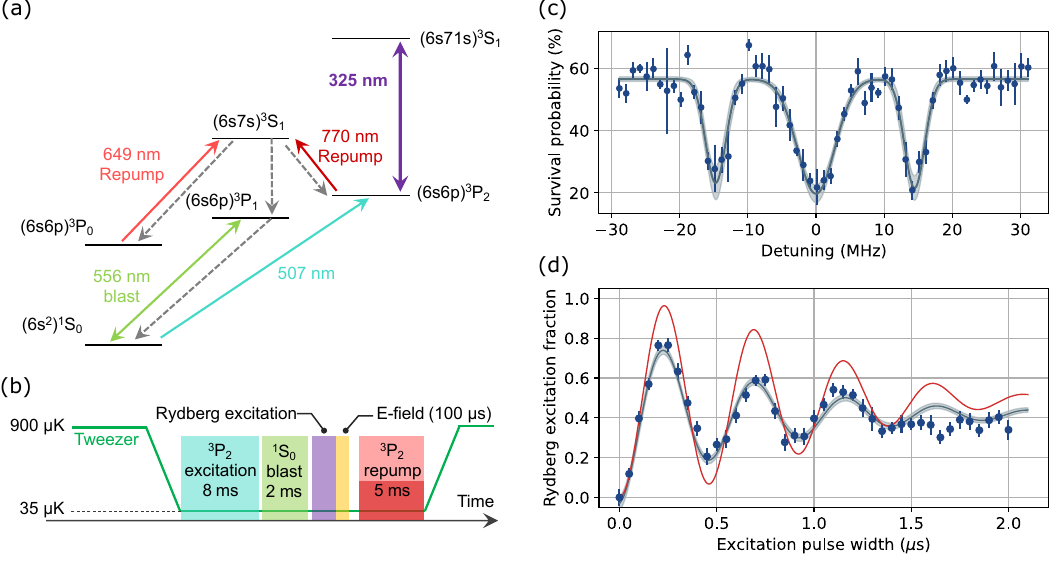}
    \caption{Rydberg excitation experiments. (a) The energy levels of $\boson$ and used transitions in this experiment. (b) Experimental sequence. After imaging the atoms, the trapping potential is reduced to \qty{35}{\mu K}, and a \qty{507}{nm} laser with a frequency chirp of $\pm\qty{6}{kHz}$ around the resonance is applied for \qty{8}{ms}. Any remaining atoms in the ground state are then removed by a resonant \qty{556}{nm} beam. Subsequently, a \qty{325}{nm} beam is used for Rydberg excitation, followed by an application of an electric field pulse of \qty{29}{V/cm}. Finally, the remaining atoms in the $\tPt$ state are repumped back to the ground state and imaged after restoring the trapping potential to its original depth. The horizontal axis is not to scale. (c) Spectrum of the $(6s71s)\tSo$ Rydberg state. The gray line is the fitting curve with a triple-peak Gaussian function. The shaded area represents $1\sigma$ confidence intervals. The three peaks correspond to the magnetic sublevels of the Rydberg state. The error bars represent the standard error of the mean. (d) Coherent Rabi oscillation signal. The gray line represents the fit using a damped sine function, yielding a Rabi frequency of \qty{2.13(3)}{MHz}. The error bars represent the standard error of the mean. The red line shows a simulated dynamics model that includes measured shot-to-shot fluctuations in both intensity and frequency. The intensity fluctuation is characterized by $\sigma_1 = \qty{8.7}{\%}$, and the frequency fluctuation by $\sigma_2 = \qty{0.3}{MHz}$. In the simulation, the relative intensity and detuning are sampled from normal distributions $\mathcal{N}(1, \sigma_1^2)$ and $\mathcal{N}(0, \sigma_2^2)$, respectively, where $\mathcal{N}(\mu,\sigma^2)$ is a normal distribution with a mean of $\mu$ and a standard deviation of $\sigma$. The averaged dynamics are computed over 10,000 iterations.}
    \label{fig:figure4}
\end{figure*}

Furthermore, we evaluate the coherence of the $\tPt$ optical clock qubits using a Ramsey sequence. Figure \ref{fig:figure3}\,(d) shows the Ramsey signal obtained by scanning the holding time between the two $\pi/2$ pulses on the $\sSz\text{--}\tPt$ transition. Fitting the data with a damped sine function yields a coherence time of \qty{0.55(7)}{ms}. The oscillation frequency of \qty{76.55(4)}{kHz} corresponds to the probe light shift induced by the \qty{507}{nm} beam. The coherence time might be currently shortened due to the spatial distribution of the differential light shift between the ground and $\tPt$ states in each tweezer potential \cite{Unnikrishnan2024-ec}, as well as the aforementioned thermal distribution, which induces slight shot-to-shot variations in the Ramsey frequency. Meanwhile, we anticipate achieving a coherence time on the order of several milliseconds for the $\tPt$ optical clock qubits in magic-wavelength traps, as recently demonstrated in a strontium system \cite{Klusener2024-nz}.

\section{Coherent Rydberg excitation of single \textsuperscript{174}Yb atoms}

We next employ the UV laser system for coherent Rydberg excitation of single $\boson$ atoms trapped in a $8\times8$ optical tweezer array. The site intervals are set to \qty{10}{\mu m} in this experiment. The time sequence of the Rydberg excitation and relevant transitions are shown in Fig.\ref{fig:figure4}\,(a) and (b). The first step is the state transfer from the ground $\sSz$ state to the $m_J=0$ of the excited metastable $\tPt$ state, as described above. For better signal-to-noise ratio in the Rydberg excitation experiments, the frequency of the \qty{507}{nm} laser is chirped within a range of $\pm\qty{6}{kHz}$ around the resonance of the $\tPt$ state for \qty{8}{ms} as an adiabatic passage. Following this, we apply a \qty{325}{nm} beam with vertical linear polarization for Rydberg excitation after irradiating a resonant \qty{556}{nm} beam to blast out the remaining atoms in the ground state. Subsequently, the Rydberg atoms are quickly ionized by applying an electric field of \qty{29}{V/cm} for \qty{100}{\mu s}. We then repump back the remaining $\tPt$ atoms to the $\sSz$ state using \qty{770}{nm} and \qty{649}{nm} pulses for \qty{5}{ms}. Finally, the atoms in the $\sSz$ state are imaged to estimate the Rydberg excitation fraction.

We first perform the spectroscopy of the $(6s71s)\tSo$ Rydberg state. In this measurement, the \qty{325}{nm} beam power and pulse width are set to \qty{15}{mW} and \qty{1}{ms}, respectively. Note that the pulse width is much longer than the Rydberg state lifetime of several tens of microseconds, and the excitation fraction can be detected as atom loss even without applying the electric field. Figure \ref{fig:figure4}\,(c) shows a typical spectrum of the Rydberg state. The three dips represent the magnetic sublevels. Using the measured $\sSz\text{--}\tPt$ frequency of \qty{590.902342562(63)}{THz} \cite{yamaguch-phd}, the transition frequency of the $\sSz$ to $(6s71s)\tSo$ Rydberg state is found to be \qty{1511.5026513(2)}{THz}. The number in parenthesis represents the fitting uncertainty, and the accuracy is limited by the wavemeter, which has an accuracy of \qty{20}{MHz}. The measured value is in good agreement with the value reported in the previous research \cite{Wilson2022-gz}.

Next, we perform coherent excitation of the $m_J=0$ of $(6s71s)\tSo$ Rydberg state. For this, we increase the \qty{325}{nm} beam power to \qty{70}{mW} and align the magnetic field direction to the beam polarization during excitation to maximize the Rabi frequency per excitation beam intensity. Figure \ref{fig:figure4}\,(d) shows a Rabi oscillation between the $\tPt$ and the Rydberg state obtained by scanning the pulse width from \qty{0}{\mu s} to \qty{2}{\mu s}. The Rabi frequency is $2\pi\times\qty{2.13(3)}{MHz}$, which is as high as the previous demonstration of high-fidelity two-qubit gate for a neutral atom tweezer array \cite{Peper2024-wc}. We expect about a two-fold further improvement of the Rabi frequency by perfecting our beam shaping and alignments.

\section{Discussion}
The observed Rabi oscillation of the Rydberg excitation decays, primarily due to shot-to-shot fluctuations in both the intensity and frequency of the excitation laser. For the former, the intensity fluctuates by approximately \qty{9}{\%} around its mean value, as measured by monitoring the pulse area. This fluctuation may arise from irregularity in pulse shape due to the lack of AOM warm up inevitable to avoid UV damage to the fiber. To mitigate this issue, we can employ a sample-and-hold technique \cite{Levine2018-ge,Madjarov2020-kz} combined with an EOM-based intensity stabilization technique \cite{Wang2020-mr}. For the latter, we estimate that the frequency fluctuates by around \qty{0.3}{MHz} even though the laser is stabilized with a filtering cavity. This limitation is due to the restricted frequency resolution of the wavemeter, which is used to lock the length of the filtering cavity. This issue can be addressed by locking the seed laser to a  stable frequency reference; for example, an optical frequency comb stabilized by an ultranarrow laser \cite{Inaba2013-dt}. We simulate the system dynamics while accounting for the aforementioned fluctuations, and the results are plotted as the red line in Fig.\ref{fig:figure4}\,(d). Although the simulated excitation fraction does not perfectly match the experimental data -- indicating the presence of other potential sources of decoherence such as the atom temperature, differential light shift between the $\tPt$ and the Rydberg states, electric field noises, and so on -- the decay of the oscillation qualitatively agrees with the observed experimental behavior.

\section{Conclusion}
In summary, we have developed a high-power UV laser system for Rydberg excitation from the $\tPt$ state of Yb atoms. By employing two-stage frequency doubling with a RFA and a home-built SHG cavity, we successfully achieved an output power exceeding \qty{800}{mW} at a wavelength of \qty{325}{nm}. To mitigate frequency noise in the MHz range, which can adversely affect Rydberg excitation with an MHz-order Rabi frequency, we incorporated a filtering cavity with a linewidth of \qty{230}{kHz}. The suppression factor for the seed laser's frequency noise was measured to be \qty{20}{dB} in the MHz range, suggesting the potential for high-fidelity Rydberg excitation. Additionally, we characterized the intensity noise at \qty{325}{nm} and found that AOM-based intensity stabilization effectively suppresses up to \qty{500}{kHz}. Moving forward, the stabilization bandwidth will be further enhanced by integrating an EOM-based stabilization method \cite{Wang2020-mr}.

As a first step to the Rydberg excitation, we demonstrated coherent manipulation on the $\sSz\text{--}\tPt$ transition of $\boson$, which also serves as the optical clock qubit with the capacity of non-destructive qubit readout. We observed Rabi oscillations between the ground state and the $m_J=0$ state of the $\tPt$ with a Rabi frequency of $2\pi\times\qty{9.5(2)}{kHz}$. Furthermore, we evaluated the coherence time of the optical qubit to be \qty{0.55(7)}{ms} using the Ramsey sequence, providing the starting point for quantum computation utilizing this qubit.

Using the laser system we developed, we conducted spectroscopy and coherent excitation of the $(6s71s)\tSo$ Rydberg state. The measured resonance frequency was in good agreement with the value reported in previous research \cite{Wilson2022-gz}. By increasing the laser power, we successfully observed Rabi oscillations between the $m_J=0$ states of the $\tPt$ and the Rydberg state, with a frequency of $2\pi\times\qty{2.13(3)}{MHz}$. The decay of the Rabi oscillation was qualitatively attributed to shot-to-shot fluctuations in both the frequency and intensity of the laser. These issues can be readily addressed using existing experimental techniques.

The high-power UV laser system with frequency noise suppression presented in this work serves as a foundational step toward fast and high-fidelity Rydberg excitation. By incorporating further laser noise suppression techniques, this system has the potential to significantly advance quantum simulation and computing in neutral atom arrays.

\section{Acknowledgement}
\label{sec:acknowledgement}
This work was supported by the Grant-in-Aid for Scientific Research of JSPS (No.\ JP17H06138, No.\ JP18H05405, No.\ JP18H05228, No.\ JP21H01014, No.\ JP21K03384, No.\ JP22K20356), JST PRESTO (No.\ JPMJPR23F5), JST CREST (Nos.\ JPMJCR1673 and JPMJCR23I3), MEXT Quantum Leap Flagship Program (MEXT Q-
LEAP) Grant No.\ JPMXS0118069021, and JST Moon-shot R\&D Grant (Nos.\ JPMJMS2268 and JPMJMS2269).
Y.N.\ acknowledges support from the JSPS (KAKENHI Grant No.\ 22KJ1949). 
N.O. acknowledges support from the JSPS (KAKENHI Grant No. 24KJ0120). 
T.K.\ acknowledges the support from the establishment of university fellowships towards the creation of science technology innovation, Grant No.\ JPMJFS2123.


%

\end{document}